\documentclass[EN]{rcftex}

\usepackage[T1]{fontenc}

\title{Perturbative calculation of the Sternheimer anti-shielding factor with Hartree-Fock atomic orbitals}       

\author{E. D\'iaz Su\'arez \toaff{a}, Carlos M. Cruz Incl\'an \toaff{b} and Augusto Gonz\'alez \toaff{c}}

\affiliations{
\aff University of Informatics Sciences, UCI; edsuarez@uci.cu, eduardo.dsuarez@gmail.com\emailto
\aff Center for Technological Applications and Nuclear Development, CEADEN; ccruz@ceaden.edu.cu
\aff Institute of Cybernetics, Mathematics and Physics, ICIMAF; agonzale@icimaf.cu}

\abstractES{Reportamos el c\'alculo del factor de anti-apantallamiento de Sternheimer, $\gamma$, a partir de la teor\'ia de perturbaciones de primer orden. 
Como base de funciones se utilizan los orbitales electr\'onicos de Hartree-Fock, expandidos en estados at\'omicos hidrogenoideos. Las dependencias $\gamma(r)$, calculadas para las corazas electr\'onicas internas de los sistemas 
$Fe^{3+}$ y $Cu^{1+}$, son reportadas aqu\'i y comparadas con valores existentes en la literatura, obtenidos por
metodolog\'ias alternativas.}          

\abstractEN{We report a calculation of the Sternheimer anti-shielding factor, $\gamma$, by means of first order perturbation theory. In quality of basis functions, we use Hartree-Fock electronic orbitals, expanded on hydrogenic atomic states. The computed $\gamma(r)$ for $Fe^{3+}$ and $Cu^{1+}$ inner electronic cores are reported and compared with literature values, obtained from alternative methodologies.}

\keyW{Electric field gradient at the atomic nucleus, Sternheimer anti-shielding factor, calculation methodologies, Hartree-Fock atomic orbital functions, Hartree-Fock $Fe^{3+}$ and $Cu^{1+}$ inner electronic core configurations.}

\begin{document}

\maketitle

\section{Introduction}

In Ref. [1], the reported $^{57}Fe$ Moessbauer spectroscopy nuclear quadrupole splitting (NQS) data, measured on $YBa_2Cu_3O_{7-x}$ ($YBCO$) samples, have been ascribed to $^{57}Fe$ occupied structural sites. In this Ref., the \mbox{oxygen} nearest neighbour disorder, around the $^{57}Fe$ occupied $Cu(1)$ crystal site in the unit cell base plane, was taken into account. The electrical field gradients (EFG) were computed according to the point charge model (PCM). On this ground, satisfactory qualitative assignment of all Moessbauer spectroscopy quadrupole splitting data were achieved for two $YBCO$ crystalline phases: the oxygen-deficient tetragonal and oxygen-rich orthorhombic ones.

The next step, in the way of getting a deeper understanding of the observed $^{57}Fe$ NQS data for this material, must rely on an improvement of the present knowledge about oxygen nearest-neighborhood disorder in the $YBCO$ unit-cell basal plane. This is related to the distribution of the electronic charge in the neighborhood of the $Cu(1)$ site.

This purpose can be reached by introducing quantum \mbox{mechanical} calculations in the framework of the Heitler and London approximation (HLA)\cite{Solid_State_book}, which still offers, for complex systems, a valid approach for EFG calculations, alternative to density functional theory (DFT). \cite{Density_funct, YBaCu_01, EFG_01, EFG_02, EFG_03, EFG_04, EFG_05, EFG_06, EFG_07, EFG_08, EFG_09}

In the HLA, it is assumed that only valence electrons are involved in chemical bonds, while the atomic inner electronic core (IEC) remains unperturbed, and follows the free-atom electronic structure. The atomic IEC polarizations effects, connected to EFG calculations, are described through the Sternheimer anti-shielding factor \cite{Sternheimer01, Sternheimer02, Sternheimer03, Sternheimer04}, $\gamma(r)$, a kind of quadrupole polarizability coeffient. This $\gamma(r)$ can be expressed as the ratio between the IEC induced EFG components at the nuclear site, and the \mbox{corresponding} electrical field gradients arising from an external point electrical charge, placed at a distance $r$ to the atomic nucleus. 

The knowledge of $\gamma(r)$ is essential to insure a successful application of the HLA for EFG calculations. However,  there exists a wide spread of the reported $\gamma_{\infty}$ $(\gamma(r)\vert_{r\rightarrow\infty})$ values \cite{Sternheimer01, Sternheimer02, Sternheimer03, Sternheimer04, Shield_Fact01, Ingalls, Shield_Fact02} for the  $Fe^{3+}$ and $Cu^{1+}$ electronic systems. On the other hand, the full $\gamma(r)$ dependence have only been reported for the $Fe$ inner electronic core. \cite{Ingalls}

Therefore, for the HLA application to NQS data \mbox{evaluation} purposes, in particular to $^{57}Fe$ and $^{63,65}Cu$ hyperfine interaction in the $YBa_2Cu_3O_{7-x}$ system, further research work is required. The present contribution addresses the perturbative calculation of $\gamma_{\infty}$ and $\gamma(r)$ with the use of Hartree-Fock (HF) electronic wave functions, expanded on a hydrogenic basis set. \cite{Augusto} This methodology is applied to the $Fe^{3+}$ (Ne = 23) and $Cu^{1+}$(Ne = 28) electronic systems.

\section{Theoretical Procedure}

\subsection{Hartree-Fock Calculations}

$Fe^{3+}$ and $Cu^{1+}$ inner core electronic states are defined as $1s^2 2s^2 2p^6 3s^2 3p^6 3d^5$ and  $1s^2 2s^2 2p^6 3p^6 3s^2 3d^{10}$, respectively . Such configurations were computed in the nonrelativistic HF Approximation following an algorithm developed earlier \cite{Augusto}, where each HF orbital is a linear combination of hydrogenic functions $\chi_{\alpha}$:

\begin{equation}\label{Ec1}
\varphi_i^0=\sum_\alpha{C_{i\alpha}\chi_\alpha},
\end{equation}

\noindent where $\alpha$ denotes $(n, l, m, m_s)$; a given hydrogenic quantum numbers set, and the hydrogenic  wave functions $\chi_\alpha$ are defined by:

\begin{equation}\label{Ec2}
\chi_{\alpha}=\psi_{nlmm_s}(r,\theta,\phi,s_z)=\frac{1}{r}P_{nl}(r)Y_{lm}(\theta,\phi)\xi_{m_s}\left(s_z\right),
\end{equation}

\noindent
where:

\begin{equation}\label{Ec3}
P_{nl}(r)=\sqrt{\frac{Z(n-l-1)!}{n^2{\left[(n+l)!\right]}^3 }} {\left(\frac{2Zr}{n}\right)}^{l+1}e^{-\frac{Zr}{n}}L_{n+l}^{2l+1}\left( \frac{2Zr}{n}\right).
\end{equation}

\noindent
In the last Eq., $L_{n+l}^{2l+1}\left( \frac{2Zr}{n}\right)$ are Laguerre polynomials, $Y_{lm}(\theta,\phi)$ are spherical harmonics \cite{Weissb}, and $\xi_{m_s}\left(s_z\right)$ are spin functions with defined projection along the direction OZ.

Restricted HF calculations are performed, leading to quantum states, $\varphi_i^0$,  which are energetically degenerated with respect to the quantum numbers $m$ and $m_s$. The resulting partial charge densities are spherically symmetric, $\rho_{n'l}=\sum_{m,m_s}{\vert\varphi_{n'lmm_s}^0\left(r,\theta,\phi,s_z \right)\vert^2}=\left(\frac{2l+1}{4\pi}\right)\left( R_{n'l} \right)^2\left(r\right)$, leading to a vanishing contribution to the EFG tensor at the nucleus.
\vspace{.3cm}

\subsection{Sternheimer anti-shielding factor calculation}\label{metod}

The Sternheimer anti-shielding factor $\gamma(r)$ is related to IEC polarization effects due to an external electric charge. 

\begin{figure}
\includegraphics[width=8cm]{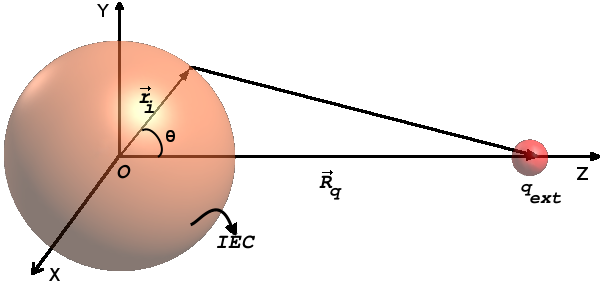}
\caption{\label{fig1} The unperturbed inner electronic core and the external point charge used in the present calculation.}
\end{figure}

Fig.\ref{fig1} schematically shows the inner core and the external point charge $q$ used in our calculation. The unperturbed inner core charge density is spherically symmetric. The coordinate system origin is fixed at the atomic nucleus (nuclear dimensions are neglected). $\vec{r_i}$ and $\vec{R_q}$ represent the $ith$ inner core electron, and the external charge q coordinate vectors, respectively. 

The external charge induces a polarization of the electronic cloud. The rate between $G_{Qz}$, the z component of the EFG due to the polarized inner electronic core, and $G_{qz}$, the z component of the EFG caused by the bare point charge, both taken at the nucleus site, is defined as the Sternheimer anti-shielding factor, according to the equation,

\begin{equation}\label{gamma_rq}
\gamma(R_q) =\frac{G_{Qz}}{G_{qz}} =\frac{{R_q}^3}{q_{ext}}\int{\delta\rho(r,\theta) \frac{ P_2(\cos\theta)}{r^{3}}dv}.
\end{equation}

Here, $\delta\rho(r,\theta)$ represents the polarization charge (deviation from radial symmetry). It is computed in first order perturbation theory from the \mbox{following} quadrupole perturbation Hamiltonian (in atomic units),

\begin{equation}\label{Ec5}
\hat{H}_Q = 
\left\{ \begin{array}{ll}
-q_{ext} \sum_i{\left({{r_i}^2}/{{R_q}^3}\right) P_2 \left(cos⁡\theta \right)}, & \mbox{for } r_i< R_q  \\
-q_{ext} \sum_i{\left({{R_q}^2}/{{r_i}^3}\right) P_2 \left(cos⁡\theta \right)}, & \mbox{for } R_q < r_i \\
\end{array} \right.
\end{equation}

\noindent
That is:

\begin{equation}\label{Ec6}
\delta\rho(r,\theta) = \sum_i^{occupied}{\delta\rho_i(r,\theta)} = \sum_i^{occupied}{2\varphi_i^{1}\varphi_i^{0}},
\end{equation}

\noindent where the $\varphi_i^{1}$ are first-order corrections to the wave functions:

\begin{equation}\label{Ec7}
\varphi_i^{1} = \sum_{j\neq i}{B_{ij}\varphi_j^{0}},
\end{equation}

\noindent
and the coefficients for non-degenerate levels, for example, are computed as $B_{ij}= -{\langle{\varphi_j^{0}}\vert{\hat H^i_Q}\vert{\varphi_i^{0}}\rangle}/{\left(E_j^{0} - E_i^{0}\right)}$. Selection rules dictate that $l_j=l_i$ or $l_i\pm 2$ (radial and angular contributions), and lead to many vanishing terms in Eq. (\ref{Ec7}). On the other hand, as  $j$ increases, $B_{ij}$ decreases.)

For the Sternheimer anti-shielding factor, we get the following explicit expression:

\begin{equation}\label{Ec8}
\gamma(R_q) = \sum_{i}^{occupied}{\gamma_i},
\end{equation}

\noindent where 
\begin{equation}\label{Ec9}
\gamma_i = \frac{R_q^3}{q_{ext}}\sum_{j\neq i}{B_{ij} \left\langle{\varphi_i^{0}}\left\vert{\frac{P_2 \left(cos⁡\theta \right)}{r^3}}\right\vert{\varphi_j^{0}}\right\rangle}.  
\end{equation}

\noindent
Radial and angular integrations in Eq. (\ref{Ec9}) can be separatedly performed. The radial integrals change sign, which leads to alternating contributions to $\gamma_i$.
\vspace{.3cm}

\section{Results and Discussion}

\subsection{The program}

A program (antish2.001), written in C++ for the GNU (g++) compiler, performs the Sternheimer anti-shielding factor estimation for a given electronic core. It uses as input data the restricted HF calculations for the given system, in particular the IEC occupied as well as the first 200 virtual orbitals. HF orbitals are expanded in a basis of 400 hydrogenic states.

In our program, the radial integrals are analytically calculated. The computation of 3J-coefficients and the diagonalization of the $\hat H^i_Q $ matrix is performed with the help of the GNU Scientific Library (GSL).
\vspace{.3cm}

\subsection{Computational and convergence details}\label{details_}

Due to $l$ quantum number selection rules, many $B_{ij}$ coefficients in Eq. (\ref{Ec7}) vanish. The convergence of the series is relatively slow. Notice that the partial radial and angular contributions to $\gamma(R_q)$ show opposite signs (see Tables II and III below). 

The number of Hartree-Fock virtual orbitals is a critical subject for convergence. In the present calculations, convergence is reached when using 180 or more virtual levels, guaranteeing relative variations lower than 0.2\%, as can be observed in Table I for $Fe^{3+}$ $\gamma(R_q=5)$ calculations.

\begin{figure}
\includegraphics[width=8cm]{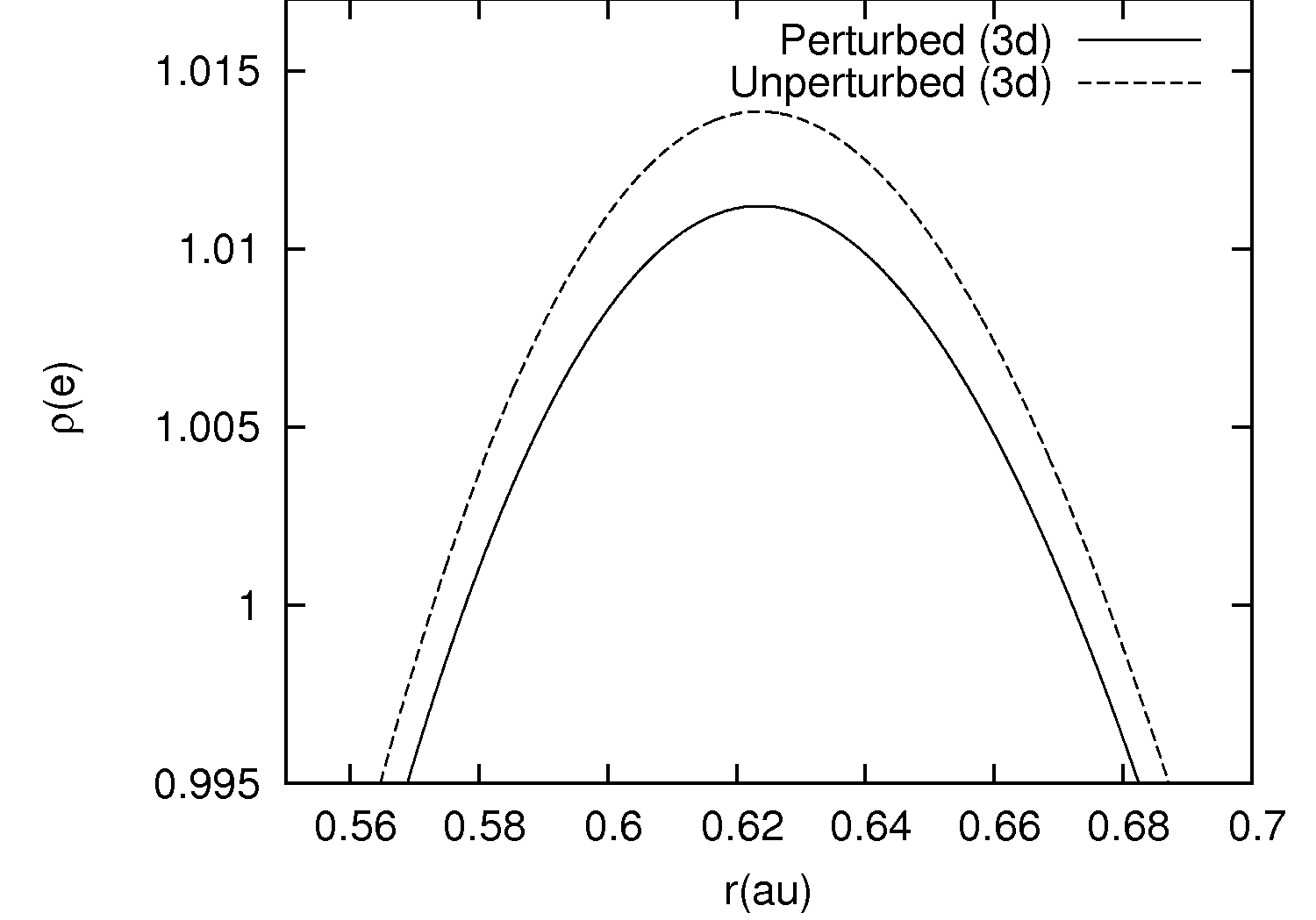}
\caption{\label{density_Cu} The IEC $3d$ orbital of $Cu^{1+}$ in the vicinity of its maximum. The quadrupole field of a charge at $R_q=5$ a.u. from the nucleus causes the perturbation.}
\end{figure}

The computed $\gamma(R_q)$ \mbox{values} are very sensitive to small induced variations of the IEC wave functions. In Fig. \ref{density_Cu}, we show both the unperturbed and perturbed $Cu^{1+}$ $3d$ orbitals in the neighborhood of their first maximum, which present a relative variation of about 0.2\%. This certainly small functional change induces a radial contribution of -5.301 to $\gamma(R_q=5)$ (see Table II below).

\begin{tablenviroment}
\caption{\label{TAB1} Convergence of $\gamma(R_q=5)$ for $Fe^{3+}$ with respect to the number of virtual orbitals
included in the calculation.}
\begin{tabular}{|c|c|}
\tabhead{2}{7.6cm} 
\hline
Number of virtual orbitals &  $\gamma(R_q)$  \\
\hiderowcolors 
 99 & -4.84778498 \\ \hline 
 130 & -7.71152215 \\ \hline
 140 & -7.71152215 \\ \hline
 150 & -7.71152215 \\ \hline
 160 & -7.71152215 \\ \hline
 170 & -7.87306174 \\ \hline
 180 & -7.81577288 \\ \hline
 190 & -7.79963097 \\ \hline
 200 & -7.79963097 \\ 
 \hline
\end{tabular}
\end{tablenviroment}

\vspace{.3cm}
\subsection{$\gamma(R_q)$ and $\gamma_\infty$ results for $Fe^{3+}$ and $Cu^{1+}$}

Partial contributions to $\gamma(R_q=5)$ for $Cu^{1+}$ and $Fe^{3+}$ are presented in Tables II and III, respectively, in which, for comparison purposes, values reported for $\gamma_\infty$ by other authors are also included. The reader may find more details in Ref. 24.

\begin{tablenviroment}
\caption{\label{TAB2} Partial contributions to $\gamma(R_q=5)$ in $Cu^{1+}$. Radial and angular contributions, as well as the $\gamma_\infty$ values reported by other authors are included.}
\begin{tabular}{|l|l|l|l|l|l|}
\tabhead{6}{7.6cm} 
          \hline
          Partial & This & Ref. & Ref. & Ref. & Ref. \\ 
          Contrib.& Work & 14   & 15   & 23   & 18   \\
          \hiderowcolors 
          \hline
          2p-p*  & -0.498 & -0.6 & -0.62 & -0.58 & - \\
          3p-p* & -6.580 & -6.4 & -7.9 & -13.01 & - \\
          3d-d* & -5.301 & -3.7 & -8.5 & -16.7  & - \\
          Total radial     &-12.379 &-10.7 & -17.02 & -30.29 & - \\
          Angular & 0.320   & 2.0  &  2.0   &  1.19  & - \\
         \hline
         $\gamma(R_q=5)$   &-12.059  &-8.7  & -15.02 & -29.10 & -17.37 \\
         \hline
\end{tabular}
\end{tablenviroment}

\begin{tablenviroment}
\caption{\label{TAB3}The same as Tab. II for $Fe^{+3}$.}
\begin{tabular}{|l|l|l|l|l|l|}
\tabhead{6}{7.6cm} 
    \hline
    Partial &This  & Ref. & Ref. & Ref. & Ref.  \\
    Contrib.&Work  & 17   & 19   & 23   & 18    \\
    \hiderowcolors 
    \hline
      2p-p*   & -0.584    &  -0.724   &  0.78  &    -0.68  &  - \\
      3p-p*   &  -6.463   &  -8.607  &  -9.98  &    -5.00  & - \\
      3d-d*   &  -1.094   &  -2.478  &  -2.39  &    -1.53  & - \\
      Total radial  &  -8.141   & -11.809  & -11.59  &    -7.21  & - \\
      Angular   &   0.341   &   0.837  &   1.0   &     1.04  & - \\
    \hline
     $\gamma(R_q=5)$  &   -7.800  & -10.9719 & -10.59  &    -6.17  & -5.244  \\
    \hline
    \end{tabular}
\raggedright{\caption*{(*) Transitions to excited orbitals with the same angular momentum.}}
\end{tablenviroment}

Notice that our results for $\gamma$ in $Cu^{1+}$ and $Fe^{3+}$  are within the range reported by other authors, which, on the other hand, show a wide spread, presumably because of the sensitivity of $\gamma$ to small variations in the polarized charge densities.

\begin{figure}
\includegraphics[width=8cm]{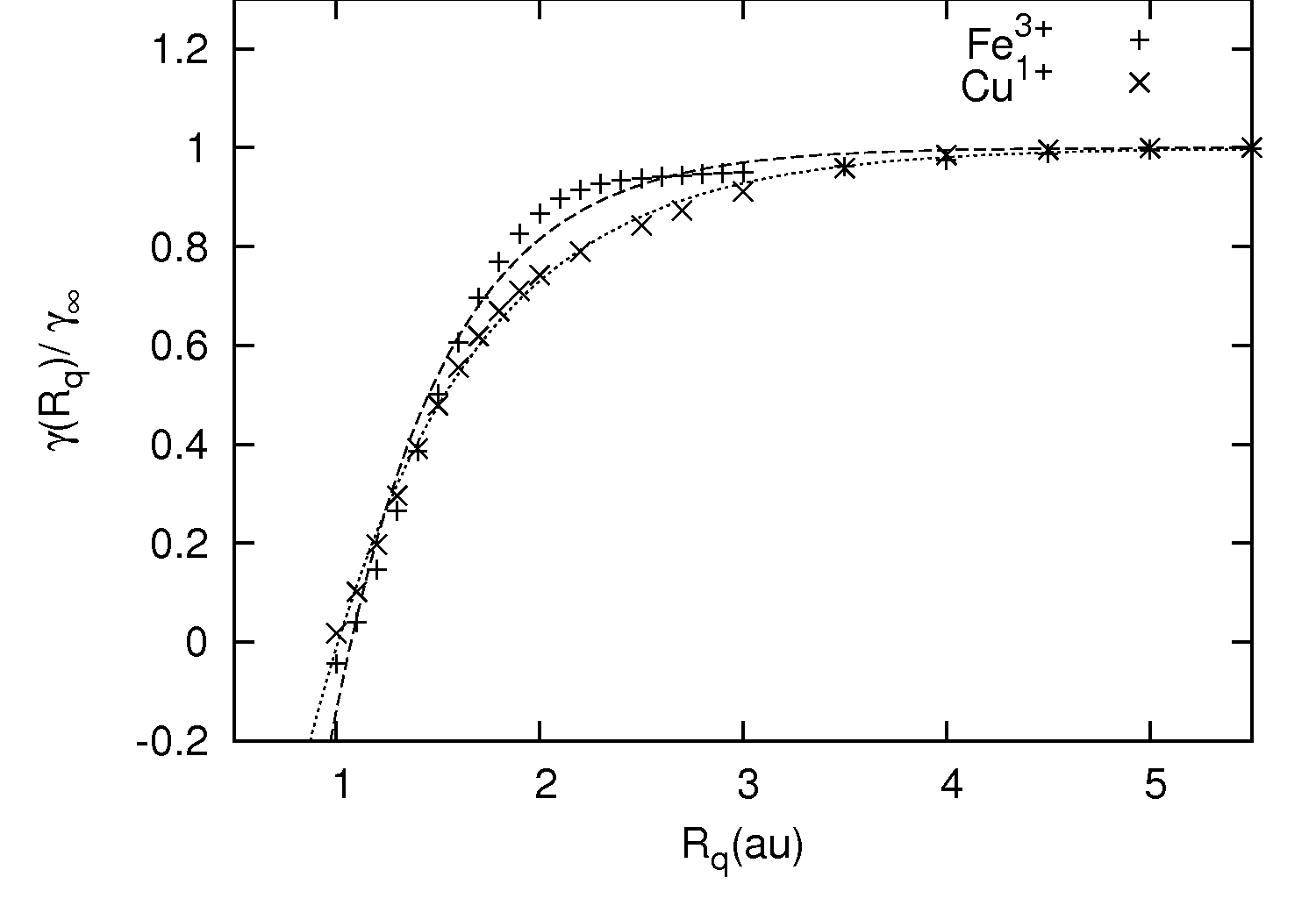}
\caption{\label{gamma_r}$Cu^{1+}$ and $Fe^{3+}$ Sternheimer anti-shielding factor dependence on $R_q$. Dotted lines are fitting curves according to $\gamma(R_q)/\gamma_\infty = 1 + \lambda *e^{(-R_q/{\rho})}$, where $\gamma_\infty(Cu^{1+})$ and $\gamma_\infty(Fe^{+3})$ are taken from Tables II and III, respectively.}

\end{figure}

We show in Fig. 3 normalized, $\gamma(R_q)/\gamma_\infty$, results for the studied $Cu^{1+}$ and $Fe^{3+}$ systems.
Notice that, in both cases, $\gamma(R_q >5) \approx \gamma_\infty$. The dotted lines in Fig. \ref{gamma_r}, represent fitting curves according to $\gamma(R_q)/\gamma_\infty = 1 + \lambda *exp(-R_q/{\rho})$, where $\lambda(Cu^{1+})=-3.797$, $\rho(Cu^{1+})=0.756$, and $\lambda(Fe^{3+})=-7.027$, $\rho(Fe^{3+})=0.550$. 

\section{Conclusions}

The Sternheimer anti-shielding factor, $\gamma(R_q)$, was successfully computed in the framework of first order perturbation theory and a hydrogenic function basis set. The HF calculations were performed in a restricted spherically symmetric approach. 

The computed $\gamma(R_q)$ dependences and $\gamma_\infty$ values for the $Cu^{1+}$ and $Fe^{3+}$ are within reasonable ranges. The calculation methodology developed in the present paper may be extended and applied to other atomic systems.

The obtained $\gamma(R_q)$ may be used in the computation of nuclear quadrupole splittings in the solid state systems. This is left for a future work. 





%
%

\end{document}